\documentclass{article}

\usepackage[table]{xcolor}
\usepackage{algorithm}
\usepackage{algorithmic}

\definecolor{commentcolor}{RGB}{110,154,155}   

\usepackage{xcolor} 

\definecolor{commentcolor}{rgb}{0.5,0.5,0.5} 

\usepackage{PRIMEarxiv}
\usepackage{multirow}
\usepackage[utf8]{inputenc} 
\usepackage[T1]{fontenc}    
\usepackage{hyperref}       
\usepackage{url}            
\usepackage{booktabs}       
\usepackage{amsfonts}       
\usepackage{nicefrac}       
\usepackage{microtype}      
\usepackage{lipsum}
\usepackage{fancyhdr}       
\usepackage{graphicx}       
\graphicspath{{media/}}     
\usepackage{amsmath}
\usepackage[flushleft]{threeparttable}
\definecolor{commentcolor}{RGB}{110,154,155}   
\newcommand{\PyComment}[1]{\ttfamily\textcolor{commentcolor}{\# #1}}  
\newcommand{\PyCode}[1]{\ttfamily\textcolor{black}{#1}} 

\title{UCM-Net: A Lightweight and Efficient Solution for Skin Lesion Segmentation using MLP and CNN

}

\author{
  Chunyu Yuan\\\\
   The Graduate Center, City University of New York\\
  \texttt{cyuan1@gradcenter.cuny.edu} \\
  \And
  Dongfang Zhao\\\\
     University of Washington\\
     \texttt{dzhao@uw.edu} \\
   \And
  Sos S. Agaian \\\\
   The Graduate Center, City University of New York\ \\
  College of Staten Island, City University of New York \\
  \texttt{sos.agaian@csi.cuny.edu} \\
}

\begin{document}
\maketitle

\begin{abstract}
 \textcolor{black}{
Skin cancer presents a formidable public health challenge, with its incidence expected to rise significantly in the coming decades. Early diagnosis is vital for effective treatment, highlighting the importance of computer-aided diagnosis systems in detecting and managing this disease. A critical task in these systems is accurately segmenting skin lesions from images, which is essential for further analysis, classification, and detection. This task is notably complex due to the diverse characteristics of lesions, such as their appearance, shape, size, color, texture, and location, compounded by image quality issues like noise, artifacts, and occlusions. While recent advancements in deep learning have shown promise in skin lesion segmentation, they often need more computational and extensive parameter requirements, limiting their practicality for mobile health applications. 
Addressing these challenges, we introduce UCM-Net, a novel, efficient, and lightweight model that synergizes the strengths of Multi-Layer Perceptrons (MLP) and Convolutional Neural Networks (CNN). The innovative UCM-Net Block diverges from conventional UNet architectures, offering significant reductions in parameter count and enhancing the model's learning efficiency. This results in robust segmentation performance while maintaining a low computational footprint. UCM-Net's performance is rigorously evaluated using the PH2, ISIC 2017, and ISIC 2018 datasets. It demonstrates its effectiveness with fewer than 50K parameters and requires less than 0.05 Giga-Operations Per Second (GLOPs). Moreover, its minimal memory requirement is just 1.19MB in CPU environment positions. It is a potential benchmark for efficiency in skin lesion segmentation, suitable for deployment in resource-constrained settings. In order to facilitate accessibility and further research in the field,the UCM-Net source code is \href{https://github.com/chunyuyuan/UCM-Net}{https://github.com/chunyuyuan/UCM-Net}.}

\end{abstract}


\section{Introduction}

Skin cancer poses a significant global health concern and stands as one of the leading cancer types worldwide. Skin cancer can be broadly categorized into two types: melanoma and non-melanoma. While melanoma accounts for only 1\% of cases, it is responsible for the majority of deaths due to its aggressive nature. In 2022, it was estimated that melanoma would account for approximately 7,650 deaths in the United States, affecting 5,080 men and 2,570 women \cite{skinstat, siegel2022cancer}. In addition, it is estimated that the United States will have 97,610 new cases of melanoma in 2023. Current statistics suggest that one in five Americans will develop skin cancer at some point in their lives, underscoring the gravity of this issue. Over the past few decades, skin cancer has emerged as a substantial public health problem, resulting in annual expenses of approximately \$ 8.1 billion in the United States alone \cite{siegel2023cancer}. \\
Skin cancer \cite{marks1995overview} is a prevalent and potentially life-threatening disease affecting millions worldwide. Among the various types of skin cancer, malignant melanoma is known for its rapid progression and high mortality rate if not detected and treated early. Early and accurate diagnosis is, therefore, critical to improving patient outcomes. Medical imaging \cite{lehmann1999survey}, particularly dermatoscopy and dermoscopy, is crucial in diagnosing skin cancer. Dermatologists and healthcare professionals rely on these imaging techniques to examine and analyze skin lesions for signs of malignancy. However, the manual interpretation of such images with a naked eye is a time-consuming and error-prone process, heavily reliant on the expertise of the examining physician \cite{nasir2018improved,barata2021explainable}.

\begin{figure}[h]
  \centering
  \includegraphics[width=0.8\linewidth]{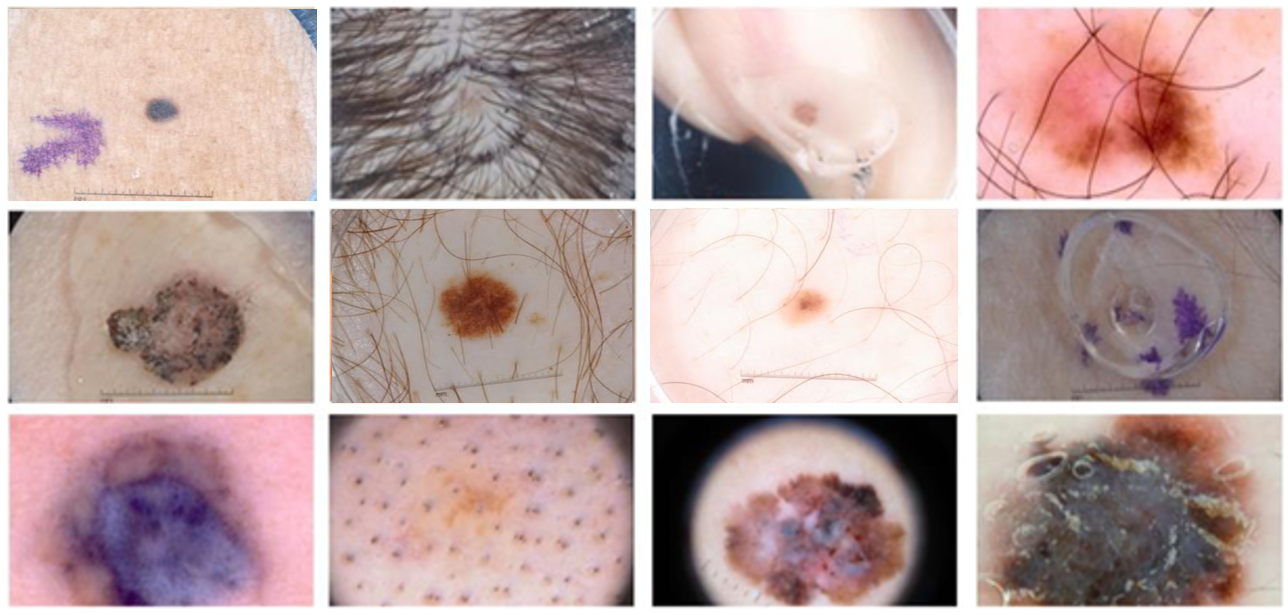}
  \caption{Complex skin lesion samples}
  \label{sample}
  
\end{figure}

To address these challenges and improve the accuracy and efficiency of skin cancer diagnosis, computer-aided tools and artificial intelligence (AI) have been leveraged in recent years\cite{sanchez2012computer,sanchez2012new,liew2023mitigation}. Skin cancer segmentation, a fundamental step in the diagnostic process, involves delineating the boundaries of skin lesions within medical images. This task is essential for quantifying lesion characteristics, monitoring changes over time, and aiding in the decision-making process for treatment. Segmenting skin lesions from images faces several key challenges \cite{hosny2023deep}: unclear boundaries where the lesion blends into surrounding skin; illumination variations that alter lesion appearance; artifacts like hair and bubbles that obscure lesion boundaries; variability in lesion size and shape; different imaging conditions and resolutions; age-related skin changes affecting texture; complex backgrounds that hinder segmentation; and differences in skin color due to race and climate. Figure \ref{sample} presents some representative samples of complex skin lesion. 

\begin{figure}[h]
  \centering
  \includegraphics[width=0.8\linewidth]{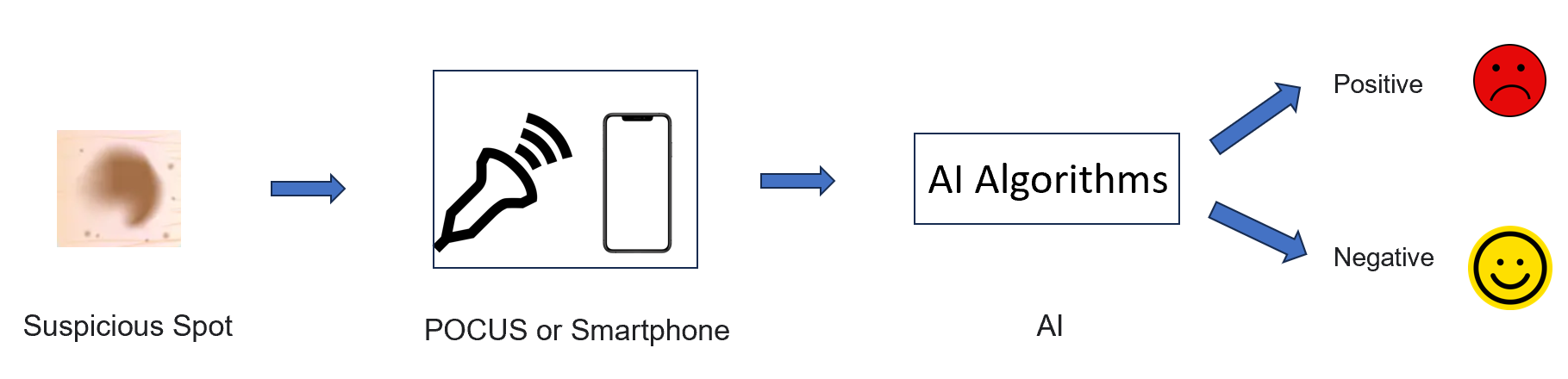}
  \caption{AI diagnose process of skin cancer detection}
  \label{process}
  
\end{figure}

Overcoming these difficulties is crucial for accurate segmentation to enable early diagnosis and treatment of malignant melanoma. Recently, a groundbreaking transformation in skin cancer segmentation has been driven by the development of advanced deep-learning algorithms \cite{frants2020dermoscopic,razzak2018deep, maier2019gentle, sharma2010automated,mirikharaji2023survey}. These AI-driven approaches have exhibited remarkable capabilities in automating the segmentation of skin lesions, significantly reducing the burden on healthcare professionals and potentially improving diagnostic accuracy. In addition, the rapid advancements in AI techniques and the widespread adoption of smart devices,such as the point-of-care ultrasound (POCUS) devices or smartphones \cite{de2019development,pocus,phonemedical},  have brought about transformative changes in the healthcare industry \cite{vashist2017point}. Figure \ref{process} briefly presents  the entire diagnose of skin cancer detection with portable devices and AI techniques. 

Patients now have greater access to medical information, remote monitoring, and personalized care, leading to increased satisfaction with their healthcare experiences. However, amidst these advancements, there are still challenges that need to be addressed.  One such challenge is the accurate and efficient segmentation of skin lesions for diagnostic purposes within limited computation hardwares and devices. Most of AI medical methods are developed based on deep-learning \cite{esteva2019guide}. The major deep-learning methods utilize expensive computation overhead and a large number of learning parameters to achieve a good prediction result. It is a challenge to embed these methods to hardware-limit devices \cite{chen2020deep, thompson2020computational}. In this study, we introduce UCM-Net, a lightweight and robust approach for skin lesion segmentation. UCM-Net leverages a novel hybrid module that combines Convolutional Neural Networks (CNN) and Multi-Layer Perceptrons (MLP) to enhance feature learning while reducing parameters. Utilizing group loss functions, our method surpasses existing machine learning-based techniques in skin lesion segmentation.

 Key contributions of UCM-Net include:

\begin{enumerate}
    \item \textbf{Hybrid }Module: We introduce the UCM-Net Block, a hybrid structure combining CNN and MLP with superior feature-learning capabilities and reduced computation and parameters.

    \item \textbf{Efficient Segmentation}: UCM-Net is developed based on UCM-Net Blocks and the base model U-Net, offering a highly efficient method for skin lesion segmentation. It is the first model with less than \textbf{50} KB parameters and less than \textbf{0.05} Giga-Operations Per Second (GLOPs). UCM-Net is \textbf{1177} times faster and has \textbf{622} times fewer parameters than U-Net. Compared to the state-of-the-art EGE-UNet, UCM-Net reduces parameter and computation costs by \textbf{1.06x} and \textbf{1.56x}.
    \textcolor{black}{ \item \textbf{Light-weight Segmentation}: Compared to published models, our model contains lower number of parameters. Besides,we further reduced the model's memory usage by engineering methods without third party library. Finally, our UCM-Net only spent 1.19MB on CPU environment. }
    \textcolor{black}{ \item \textbf{New Group Loss Function}: In the training parts, we present  a new group loss function with the output loss and internal stage losses. The new designed function can improve UCM-Net's learning ability.
    \item \textbf{Improved Segmentation}: UCM-Net's segmentation performance is evaluated using mean Intersection over Union (mIoU) and mean Dice similarity score (mDice). On the PH2, Isic2017 and Isic2018 datasets, UCM-Net enhances the baseline U-Net model by an average of \textbf{2.53\% }in mIoU and \textbf{1.50\%} in mDice. Notably, UCM-Net outperforms the state-of-the-art EGE-UNet on Ph2, ISIC 2017 and ISIC 2018 datasets, with respective mean IoU scores of \textbf{89.62\% (UCM-Net)} vs 88.39\% , \textbf{80.71\% (UCM-Net)} vs 80.22\% and \textbf{81.26\% (UCM-Net)} vs 81.16\%.}

\end{enumerate}


\begin{figure}[h]
  \centering
  \includegraphics[width=0.8\linewidth]{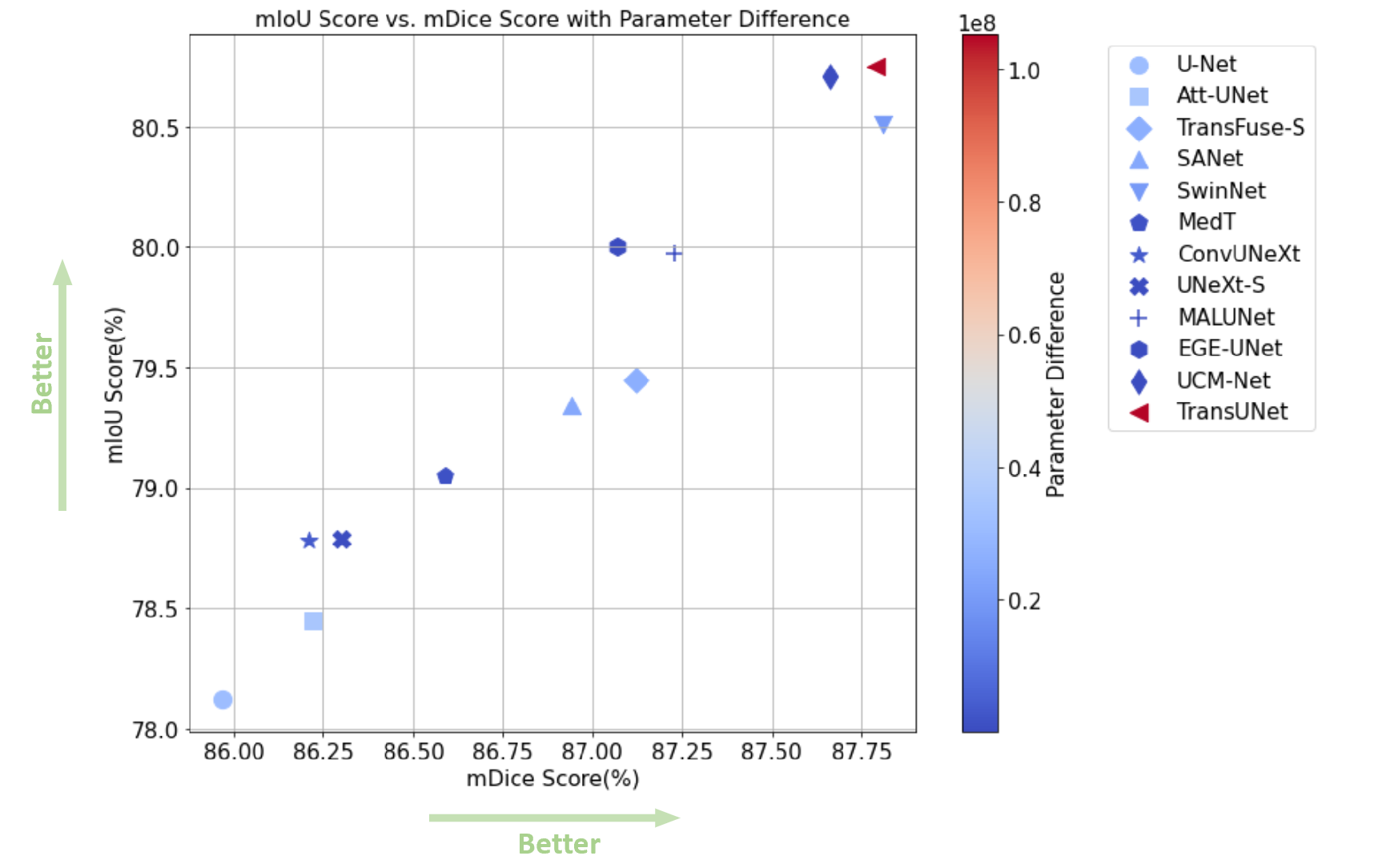}
  \caption{This figure shows the visualization of comparative experimental
results on the ISIC2017 dataset. The X-axis represents mDice score (higher is better), while Y-axis represents mIoU (higher is better). The color
depth represents the number of parameters (blue is better). }
  
\end{figure}

\section{Related works}

\textbf{AI Method Categories and Applications} AI-driven approaches for biomedical images can be broadly
classified as supervised learning methods, semi-supervised learning methods, and unsupervised learning methods
\cite{seo2020machine, serte2022deep,litjens2017survey}. Supervised learning is a solution with labeled image data, expecting to develop predictive capability. The
labeled image data can be the previous patients’ diagnosed results, such as computed tomography(CT), with Clinicians’
analysis. With the labeled data, AI-driven solutions can be developed and performed against the ground
truth results. Supervised learning solutions are widely applied to disease classification, tumor location detection, and
tumor segmentation \cite{aljuaid2022survey}. Relatively, unsupervised learning is a discovering process, diving into unlabeled data to capture
hidden information. Unsupervised learning solutions derive insights directly from unlabeled medical data without inadequate or biased human supervisions and can be used for information compression, dimensional reduction, super resolution for medical image and sequence data detection and analysis such as protein, DNA and RNA \cite{raza2021tour}. In recent years, semi-supervised learning is becoming popular, which utilizes a large number of unlabeled data in
conjunction with the limited amount of labeled data to train higher-performing models. Semi-supervised learning solutions can be also applied into disease classification and medical segmentation \cite{liu2020semi,jiao2022learning}. 

 \textcolor{black}{
\textbf{TinyML for healthcare }
Integrating the potential of TinyML into healthcare, particularly for tasks such as lesion segmentation, offers an exciting avenue for future research and practical applications. TinyML, which refers to the deployment of machine learning models on low-power, miniature hardware, presents a unique opportunity to bring advanced analytical capabilities directly to the point of care. This approach could democratize access to sophisticated medical image analysis technologies, enabling real-time, on-device processing in environments where traditional computing resources are scarce or in mobile healthcare applications. For instance, applying TinyML for lesion segmentation could facilitate immediate diagnostic insights during patient examinations or in remote areas, significantly reducing the time and infrastructure typically required for such analyses. The integration of TinyML into healthcare devices promises to enhance diagnostic processes, improve patient outcomes, and extend the reach of advanced medical technologies to underserved regions.To further enhance the efficiency and feasibility of deploying TinyML in such critical applications, researchers are exploring innovative approaches like hyper-structure optimization\cite{kim2019efficient} and the use of quantitative methods such as binary neural networks\cite{yuan2023comprehensive}. Hyper-structure optimization aims to minimize the number of parameters without compromising the model's performance, thus ensuring that the models are lightweight yet effective enough for deployment on tiny devices.  As we explore the optimization and application of UCM-Net in this context, we acknowledge the pioneering work and insights offered by Partha. in their exploration of TinyML's potential in healthcare \cite{ray2022review}, which serves as a survey foundational reference for our proposed direction.}

\textbf{Supervised Methods of Segmentation } As the technique evolves and develops, the solution of AI for medical image segmentation is from purely applying a convolution neural network(CNN) such as U-Net and Att-UNet \cite{oktay2018attention} to a hybrid structure method like TransUNet \cite{chen2021transunet}, TransFuse \cite{zhang2021transfuse} and SANet \cite{cao2022swin}. U-Net is a earest CNN solution on biomedical image segmentation, which replaces pooling operators by upsampling operators. Att-UNet is developed on the top of U-Net adding attention structures. \textcolor{black}{TransUNet merits both Transformers and U-Net for medical image segmentation.} TransFuse is a novel approach that combines Transformers and CNNs with late fusion for medical image segmentation, achieving a strong performance while maintaining high efficiency, with potential applications in various medical-related tasks. SANet \cite{wei2021shallow}, the Shallow Attention Network, addresses challenges in polyp segmentation by mitigating color inconsistencies, preserving small polyps through shallow attention modules, and balancing pixel distributions, achieving remarkable performance improvements. Swin-UNet is a UNet-like pure Transformer for medical image segmentation that proposed shifted windows as the encoder to extract context features, and a transformer-based decoder with patch expanding layer performs the up-sampling operation to restore the spatial resolution of the feature maps. MedT \cite{valanarasu2021medical} is also a transformer-based network architecture, a gated axial-attention model that introduces an additional control mechanism in the self-attention module.  ConvUNeXt \cite{han2022convunext}, an efficient model inspired by ConvNeXts \cite{liu2022convnet} and based on the classic UNet architecture, achieving excellent medical image segmentation results with a significantly reduced parameter count while incorporating features such as large convolution kernels, depth-wise separable convolution, residual connections, and a lightweight attention mechanism. UNeXt \cite{valanarasu2022unext} is introduced as an efficient Convolutional multilayer perceptron (MLP) based network that reduces parameters and computational complexity while achieving superior segmentation performance through tokenized MLP blocks and channel shifting, making it suitable for point-of-care applications. MALUNet \cite{ruan2022malunet} and its extended version EGE-UNet  \cite{ruan2023ege} develop new attention modules to significantly reduce parameters and computational complexity while achieving powerful skin lesion segmentation performance, making it highly suitable for resource-constrained clinical environments.\\

\begin{figure}[!h]
  \centering
  \includegraphics[width=0.85\linewidth]{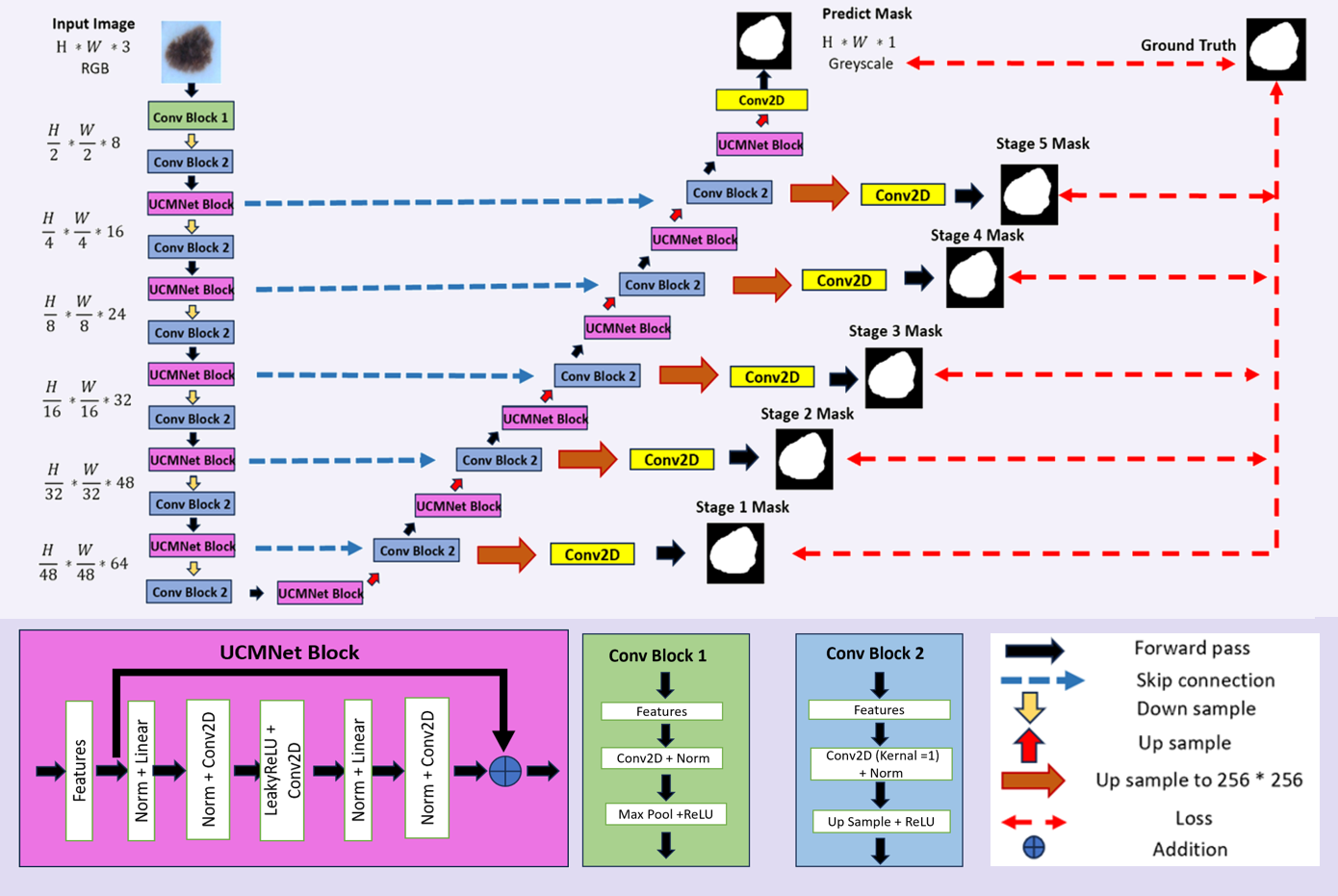}
  \caption{UCM-Net Structure}
  \label{UCM-Net Structure}
  
\end{figure}

\section{UCM-Net}
\label{sec:ucm}

\textbf{Network Design} Figure \ref{UCM-Net Structure} provides a comprehensive view of the structural framework of UCM-Net, an advanced architecture that showcases a distinctive U-Shape design. Our design is developed from U-Net. UCM-Net includes a down-sampling encoder and an up-sampling decoder, resulting in a high-powered network for skin lesion segmentation.
The entirety of the network encompasses six stages of encoder-decoder units, each equipped with channel capacities of \{8, 16, 24, 32, 48, 64\}. Within each stage, we leverage a convolutional block alongside our novel UCMNet block, facilitating the extraction and acquisition of essential features. In the convolutional block, we opt for a kernel size of 1, a choice that serves to further curtail the parameter count. Our innovative UCMNet introduces a hybrid structure module, wherein an amalgamation of a Multi-Layer Perceptron (MLP) linear component and Convolutional Neural Network (CNN) is employed, bolstered by the inclusion of skip connections. This strategic amalgamation fortifies the network's prowess in feature acquisition and learning capabilities.

 \textcolor{black}{\textbf{Convolution Block} In our designing, we contain two different convolution blocks. Convolution Block 1 utilizes a kernel size of 3×3, which is commonly employed for its effectiveness in capturing spatial relationships within the input features. This size is particularly advantageous in the network's initial layers, where preserving the spatial integrity of feature maps is essential for decoding complex input patterns.
Convolution Block 2 is characterized by a 1×1 kernel size, drastically reducing the number of learnable parameters and computational load\cite{habib2022optimization}. The smaller kernel size serves to linearly combine features across channels, allowing for dimensionality reduction and feature recombination without spatial aggregation, thus enhancing the network's efficiency.
}

\begin{algorithm}[!h]
 \PyComment{Input: \PyCode{\textbf{X}},the feature map with shape [Batch(B), Channel(C), Height(H), Width(W)]}  \\
    \PyComment{Output: \PyCode{\textbf{Out}},the feature map with shape [B, Height*Width(N),C]} \\
    \PyComment{Operator: \PyCode{\textbf{Conv}}, 2D Convolution \quad \PyCode{\textbf{LN}}, LayerNorm \quad \PyCode{\textbf{BN}}, BatchNorm, \quad  \PyCode{\textbf{Linear}}, Linear Transformation \quad \PyCode{\textbf{Leaky}}, Leaky RelU }\\
    \PyComment{UCM-Net Block Processing Pipeline} \\  
    \PyCode{B, C, H, W = X.shape()} \\
    \PyComment{Transform Feature from [B,C,H,W] to [B,H*W,C]} \\    
    \PyCode{X = X.flatten(2).trnaspose(1,2)} \\
    \PyComment{Copy feature for later residual addition} \\
    \PyCode{X1 = copy(X)} \\
   \PyCode{X = Linear(LN(X))} \\
   \PyCode{B, N, C = X.shape()} \\
   \PyComment{Transform Feature from [B,H*W,C] to [B,C,H,W]} \\  
   \PyCode{X = X.transpose(1,2).view(B,C,H,W)} \\
   \PyCode{X = Conv(LN(X))} \\
   \PyCode{X = Conv(Leaky(X))} \\
    \PyComment{Transform Feature from [B,C,H,W] to [B,H*W,C]} \\    
    \PyCode{X = X.flatten(2).trnaspose(1,2)} \\
    \PyCode{X = Linear(BN(X))} \\
   \PyComment{Transform Feature from [B,H*W,C] to [B,C,H,W]} \\  
   \PyCode{X = X.transpose(1,2).view(B,C,H,W)} \\
    \PyCode{X = Conv(LN(X))} \\
    \PyComment{Transform Feature from [B,C,H,W] to [B,H*W,C]} \\    
    \PyCode{X = X.flatten(2).trnaspose(1,2)} \\
    \PyComment{Output with residual addition}\\
      \PyCode{Out = X + X1} \\

\caption{PyTorch-style pseudocode for UCM-Net Block}
\label{algo:UCMBLOCK-algo}

\end{algorithm}
 \textcolor{black}{
\textbf{UCM-Net Block} The UCM-Net Block exemplifies a sophisticated approach to combining Convolutional Neural Networks (CNNs) with Multilayer Perceptrons (MLPs) for robust feature learning. This hybrid model benefits from the spatial feature extraction capabilities of CNNs and the pattern recognition strengths of MLPs through a series of strategically designed operations. Initially, the input feature map undergoes reshaping to cater to the distinct structural requirements of CNNs and MLPs—transitioning from a four-dimensional tensor for CNN processing to a three-dimensional tensor for MLP operations. Following this, the block applies a sequence of linear and convolutional transformations, each accompanied by normalization and activation functions. This includes the strategic use of different kernel sizes in the convolutional stages to balance computational efficiency with the network's learning capacity. Furthermore, a residual connection is integrated, enhancing information flow through the network and supporting the construction of deeper architectures by mitigating the vanishing gradient problem. Through these mechanisms, UCM-Net adeptly captures and processes complex patterns in image data, demonstrating its potential for advanced image analysis applications. The pseudocode of UCM-Net Block \ref{algo:UCMBLOCK-algo} presents our defined sequence of operations, which is how we combine CNN with MLP(Linear transformation operation) for feature learning. }


 \textcolor{black}{
\textbf{Loss functions:} In our solution, we designed a group loss function similar to those used in TransFuse \cite{zhang2021transfuse} and EGE-UNet \cite{ruan2023ege}. However, the base loss function for each stage is distinct. The base loss function from TransFuse and EGE-UNet is calculated using binary cross-entropy (BCE) \eqref{eq1} and Dice loss (Dice) \eqref{eq2} components. Our proposed base loss function is calculated from BCE components and squared-Dice loss (SDice) components, to calculate the loss from the scaled layer masks in different stages compared with the ground truth masks. Equations \eqref{eq1} and \eqref{eq2} present the stage loss in different layers and the output loss in the output layer, which are calculated using binary cross-entropy (BCE) and Dice loss (Dice) components, respectively.}
 \textcolor{black}{
\begin{flalign}
& \hspace{20mm}\text{BCE} = -\frac{1}{N} \sum_{i=1}^{N} \left[ y_i \log(p_i) + (1 - y_i) \log(1 - p_i) \right], & \label{eq1}
\end{flalign} }
 \textcolor{black}{
where $N$ is the total number of pixels (for image segmentation) or elements (for other tasks), $y_i$ is the ground truth value, and $p_i$ is the predicted probability for the $i$-th element.}
 \textcolor{black}{
\begin{flalign}
& \hspace{20mm}\text{Dice\_Loss} = 1 - \frac{2 \times \sum_{i=1}^{N} (p_i \cdot y_i) + \text{smooth}}{\sum_{i=1}^{N} p_i + \sum_{i=1}^{N} y_i + \text{smooth}}, & \label{eq2}
\end{flalign}
where $\text{smooth}$ is a small constant added to improve numerical stability.}
 \textcolor{black}{
\begin{flalign}
& \hspace{20mm}\text{Squared\_Dice\_Loss} = 1 - \frac{2 \times \left(\sum_{i=1}^{N} (p_i \cdot y_i)\right)^2 + \text{smooth}}{\left(\sum_{i=1}^{N} p_i\right)^2 + \left(\sum_{i=1}^{N} y_i\right)^2 + \text{smooth}}, & \label{eq3}
\end{flalign}
which represents an enhancement over the standard Dice loss by emphasizing the squared terms of intersections and unions.}
 \textcolor{black}{
\begin{flalign}
& \hspace{20mm}\text{Base\_loss1} = \text{BCE} + \text{Dice\_Loss}, & \label{eq4}
\end{flalign}
\begin{flalign}
& \hspace{20mm}\text{Base\_loss2} = \text{BCE} + \text{Squared\_Dice\_Loss}, & \label{eq5}
\end{flalign}}
 \textcolor{black}{
Equations \eqref{eq1} and \eqref{eq3} define the base loss functions \eqref{eq5} for our proposed model, incorporating the Dice loss, and squared-Dice loss components. $\lambda_i$ is the weight for different stages. In this paper, we set $\lambda_i$ to 0.1, 0.2, 0.3, 0.4, and 0.5 based on the i-th stage, as illustrated in Figure 4. Equations \eqref{eq8} is our proposed group loss function that calculates the loss 
from the scaled layer masks in different stages with ground truth masks. Equation  and  present the stage loss in different stage layer and output loss in the output layer.}
 \textcolor{black}{
\begin{flalign}
& \hspace{20mm}Loss_{Stage} = \text{BCE}(StagePred, Target) + \text{Squared\_Dice\_Loss}(StagePred, Target), & \label{eq6}
\end{flalign}
\begin{flalign}
& \hspace{20mm}Loss_{Output} = \text{BCE}(OutputPred, Target) + \text{Squared\_Dice\_Loss}(OutputPred, Target), & \label{eq7}
\end{flalign}
\begin{flalign}
& \hspace{20mm}Group\_Loss = Loss_{Output} + \sum_{i=1}^{5} \lambda_i \times Loss_{Stage_i} & \label{eq8}
\end{flalign}}




\section{Experiments and Results}
\subsection{Experiments Setting}
\textbf{Datasets}  \textcolor{black}{To evaluate the efficiency and performance of model with other published models, we pick the three public skin segementation datasets from PH2, International Skin Imaging Collaboration, namely ISIC2017 \cite{isic2017,berseth2017isic}
and ISIC2018 \cite{isic2018,codella2019skin}. The PH2 \cite{mendoncca2013ph} is a dermoscopic image database which 400 images of contains manual segmentation and the clinical diagnosis. The ISIC2017 dataset comprises 2150 dermoscopy images, and ISIC2018 includes 2694 images. }We noted that earlier studies\cite{ruan2023ege, ruan2022malunet} have already presented a dataset version with a pre-established train-test partition, maintaining a 7:3 ratio. In our experimental setup, we opted to utilize the previously published dataset version.

\textbf{Implementation Details} Our UCM-Net is implemented with Pytorch \cite{paszke2019pytorch} framework. All experiments are conducted on the instance node at Lambda \cite{lambda} that has a single NVIDIA RTX A6000 GPU (24 GB), 14vCPUs, 46 GiB RAM and 512 GiB SSD.  The images are normalized and resized to 256$\times$256. Simple data augmentations are applied, including horizontal flipping, vertical flipping, and random rotation. We noticed the prior studies \cite{ruan2023ege, ruan2022malunet} applied initial image processing with the calculated mean and standard deviation (std) values of the whole train and test datasets separately. While
this approach can potentially enhance their models' training and testing performance, the outcomes are notably
influenced by the computed mean and std values. Additionally, if the test dataset's context information is unknown, this operation can render the trained model less practical. In our experiment, we don't calculate the mean and std values based on the train and test datasets. Besides, TransFuse-S and SwinNet require the pre-train models with the specified input image size in their encoding stage.  \textcolor{black}{To enable fair benchmark testing, we follow the image input size for the TransFuse-S \cite{zhang2021transfuse, transfusecode},TransUNet\cite{chen2021transunet,transunetcode}  and SwinNet \cite{cao2022swin,swinnetcode} to 192$\times$256, 224$\times$224 and 224$\times$224, Correspondingly. }For the optimizer, we select AdamW \cite{loshchilov2018decoupled} initialized with a learning rate of 0.001 and a weight decay of 0.01. The CosineAnnealingLR \cite{loshchilov2016sgdr} is Utilized as the scheduler with a maximum number of iterations of 50 and a minimum learning rate of 1e-5. 
A total of 300 epochs are trained with
a training batch size of 8 and a testing batch size of 1.

\textbf{Evaluate Metrics} To assess the predictive performance of our methods, we employ mean Intersection over Union (mIoU) and mean Dice similarity score (mDice) as evaluation metrics. It's worth noting that previous studies \cite{ruan2023ege, ruan2022malunet} and \cite{valanarasu2022unext} have employed distinct calculation methods for mIoU and mDice. To comprehensively compare the performance predictions, our experiments include the presentation of mIoU, mDice, mIoU*, and mDice* results. These results are calculated using the following equations:
\begin{equation}
\text{IoU} = \frac{\text{intersection}}{\text{union}}
\end{equation}
\begin{equation}
\text{Dice} = \frac{2 \times \text{intersection} + \text{smoooth}}{\text{sum of pixels in prediction} + \text{sum of pixels in ground truth}+\text{smoooth}}
\end{equation}

where intersection represents the number of pixels that are common between the predicted output and the ground truth,  $\text{smooth}$ is a small constant added to improve numerical stability and union represents the total number of pixels in both the predicted output and the ground truth.
\begin{equation}
\text{mIoU} = \frac{1}{N} \sum_{i=1}^{N} \text{IoU}_i 
\end{equation}
\begin{equation}
\text{mDice} = \frac{1}{N} \sum_{i=1}^{N} \text{Dice}_i 
\end{equation}

where N is the number of images, \( \text{IoU}_i \) represents the IoU score for image i and \( \text{Dice}_i \) represents the Dice score for image i. 
\begin{equation}
\text{IoU*} = \frac{\text{TP}}{\text{TP} + \text{FP} + \text{FN}}
\end{equation}

\begin{equation}
\text{Dice*} = \frac{2 \times \text{TP}}{2 \times \text{TP} + \text{FP} + \text{FN}}
\end{equation}

where TP represents the number of true positive pixels, FP represents the number of false positive pixels and FN represents the number of false negative pixels. 
\begin{equation}
\text{mIoU*} = \frac{\text{TP}_{sum}}{\text{TP}_{sum} + \text{FP}_{sum} + \text{FN}_{sum}}
\end{equation}

\begin{equation}
\text{mDice*} =  \frac{2 \times  \text{TP}_{sum}}{2 \times  \text{TP}_{sum} + \text{FP}_{sum} + \text{FN}_{sum}}
\end{equation}

where \( \text{TP}_{sum} \) represents the total number of true positive pixels for images, \( \text{FP}_{sum} \) represents the total number of false positive pixels for images and \(\text{FN}_{sum} \) represents the total number of false negative pixels for images.

In our benchmark experiments, we evaluate our method's performance and compare the results among other published efficient models'. To ensure a fair comparison, we perform three sets of experiments for each method and subsequently present the mean and std of the prediction outcomes across each dataset.
\subsection{Performance Comparisons}
\begin{table}[ht!]
 \begin{threeparttable}
 \caption{Comparative prediction results on the  PH2 dataset}
  \centering
  \renewcommand{\arraystretch}{1.5} 

  \begin{tabular}{l|l|l|ll|ll}
    \toprule
   
    Dataset     & Models &Year &mIoU(\%)↑ &mDice(\%)↑     &mIoU*(\%)↑ &mDice*(\%)↑  \\
    \midrule

\multirow{ 11}{*}{PH2}

   & TransUnet$^\ast$ \cite{chen2021transunet,transunetcode} &2021&90.77 ± 0.030
 & 95.09 ± 0.014
  & 91.89 ± 0.148
& 95.77 ± 0.080
  \\
   & SANet$^\ast$ \cite{wei2021shallow,sanetcode} &2021&89.99 ± 0.349
 &  94.62 ± 0.206

  & 91.06 ± 0.481
&95.32 ± 0.264   \\

   & SwinNet$^\ast$ \cite{cao2022swin,swinnetcode} &2021&88.92 ± 0.493
 & 94.03 ± 0.289
  & 89.96 ± 0.388
& 94.71 ± 0.215
  \\
  &  TransFuse-S$^\ast$ \cite{zhang2021transfuse, transfusecode}&2021&91.15 ± 0.131 & 95.30 ± 0.078 
  &92.16 ± 0.113 &95.92 ± 0.061   \\
   \cline{2-7}
&   U-Net \cite{ronneberger2015u,unetcode}&2015 &89.09 ± 0.224
 & 94.13 ± 0.112 
  & 90.08 ± 0.545
&94.78 ± 0.302
  \\
   & Att-UNet \cite{attunetcode, oktay2018attention} &2018&89.01 ± 0.345
 &94.06 ± 0.222
  &89.82 ± 0.371

&94.64 ± 0.206

 \\

    &   MedT \cite{valanarasu2021medical,medtcode}&2021& 87.77 ± 0.099
 & 93.35 ± 0.063 
  & 88.53 ± 0.338
& 93.92 ± 0.190
    \\

  & ConvUNeXt \cite{han2022convunext, convunextcode} &2022&89.46 ± 0.056
 & 94.33 ± 0.035
  & 90.39 ± 0.172
& 94.95 ± 0.095
  \\
 &  UNeXt-S \cite{valanarasu2022unext,unextcode}&2022&89.03 ± 0.246  &94.09 ± 0.163   &89.689 ± 0.269  &94.56 ± 0.149   \\
 & MALUNet \cite{ruan2022malunet,malunetcode}&2022&87.75 ± 0.378 & 93.35 ± 0.214&88.19 ± 0.329  &93.73 ± 0.186  \\

  &    EGE-UNet \cite{ruan2023ege,egeunetcode}&2023&88.39 ± 0.303&93.73 ± 0.189
 &88.93 ± 0.802
  &94.14 ± 0.450 \\

 &    UCM-Net (ours) &2023& 89.62 ± 0.221
  &  {94.42 ± 0.124}
  &90.51 ± 0.211   &94.96 ± 0.116
 \\
    \bottomrule
  \end{tabular}
   \begin{tablenotes}
   
      \item $^\ast$: this method needs the pre-train model on training. 
    \end{tablenotes}
 \label{table1}
   \end{threeparttable}
\end{table}

\begin{table}[ht!]
 \begin{threeparttable}
 \caption{Comparative prediction results on the ISIC2017 dataset}
  \centering
  \renewcommand{\arraystretch}{1.5} 

  \begin{tabular}{l|l|l|ll|ll}
    \toprule
   
    Dataset     & Models &Year &mIoU(\%)↑ &mDice(\%)↑     &mIoU*(\%)↑ &mDice*(\%)↑  \\
    \midrule
\multirow{ 11}{*}{isic2017}

 & TransUnet$^\ast$ \cite{chen2021transunet,transunetcode} &2021&81.12 ± 0.152
 & 88.38 ± 0.157
  & 80.02 ± 0.011
& 88.90 ± 0.113
  \\

   & SANet$^\ast$\cite{wei2021shallow,sanetcode}&2021&79.34 ± 0.023
 &  87.04 ± 0.155
  & 79.34 ± 0.102
& 88.15 ± 0.060
   \\

 & SwinNet$^\ast$ \cite{cao2022swin,swinnetcode} &2021&80.51 ± 0.102
 &87.81 ± 0.231
  & 80.48 ± 0.134
& 89.19 ± 0.073
   \\

  &  TransFuse-S$^\ast$ \cite{zhang2021transfuse, transfusecode}&2021& 80.73 ± 0.312  & 88.03 ± 0.111   &79.87 ± 0.109 & 88.81 ± 0.058   \\
   \cline{2-7}
&   U-Net \cite{ronneberger2015u,unetcode} &2015&78.12 ± 0.175
 &  85.97 ± 0.196
  & 76.42 ± 0.381
& 86.63 ± 0.245
  \\
  
   & Att-UNet \cite{attunetcode, oktay2018attention}&2018 &78.45 ± 0.113
 &  86.22 ± 0.124
  & 77.14 ± 0.097
& 87.10 ± 0.062
  \\

    &   MedT \cite{valanarasu2021medical,medtcode}  &2021&79.05 ± 0.231
 &  86.59 ± 0.125
  & 77.61 ± 0.121& 87.40 ± 0.401
   \\

  & ConvUNeXt \cite{han2022convunext, convunextcode} &2022&78.78 ± 0.362
 &  86.21 ± 0.267
  & 76.98 ± 0.490
& 86.99 ± 0.313
  \\
 
 &  UNeXt-S \cite{valanarasu2022unext,unextcode} &2022& 78.79 ± 0.234
  & 86.30 ± 0.140 &77.48 ± 0.466&87.31  ± 0.296
 \\
 
 & MALUNet \cite{ruan2022malunet,malunetcode}&2022& 79.97 ± 0.389
& 87.23 ± 0.345
 &79.11 ± 0.345
&88.34 ± 0.215
 \\
  
  &   EGE-UNet \cite{ruan2023ege,egeunetcode}&2023& 80.22 ± 0.237

  &87.24 ± 0.149
&79.71 ± 0.411
&88.71 ± 0.255 \\
  
 &    UCM-Net (ours) &2023& 80.71 ± 0.345  &  {87.66 ± 0.221}
  & {79.29 ± 0.188} &{88.45 ± 0.117}
 \\
 
    \bottomrule
  \end{tabular}
   \begin{tablenotes}
   
      \item $^\ast$: this method needs the pre-train model on training. 
    \end{tablenotes}
 \label{table2}
   \end{threeparttable}
\end{table}

\begin{table}[ht!]
 \begin{threeparttable}
 \caption{Comparative prediction results on the  ISIC2018 }
  \centering
  \renewcommand{\arraystretch}{1.5} 

  \begin{tabular}{l|l|l|ll|ll}
    \toprule
   
    Dataset     & Models &Year &mIoU(\%)↑ &mDice(\%)↑     &mIoU*(\%)↑ &mDice*(\%)↑  \\
    \midrule

\multirow{ 11}{*}{isic2018} 
   & TransUnet$^\ast$ \cite{chen2021transunet,transunetcode} &2021&82.00 ± 0.152
 & 89.11 ± 0.031
  &81.68 ± 0.511
& 89.92 ± 0.010
  \\
   & SANet$^\ast$ \cite{wei2021shallow,sanetcode} &2021&80.37 ± 0.124
 &  87.87 ± 0.114

  & 79.39 ± 0.135
& 88.51 ± 0.084   \\

   & SwinNet$^\ast$ \cite{cao2022swin,swinnetcode} &2021&81.41 ± 0.069
 &  88.58 ± 0.019
  & 80.72 ± 0.069
& 89.33 ± 0.042
  \\
  &  TransFuse-S$^\ast$ \cite{zhang2021transfuse, transfusecode}&2021&81.69 ±  0.128 &  88.82 ± 0.134
  & 81.29 ±  0.084& 89.68  ± 0.070  \\

   \cline{2-7}
&   U-Net \cite{ronneberger2015u,unetcode}&2015 &79.86 ±  0.075
 &  87.57 ± 0.085
  & 78.27 ±  0.300
& 87.81 ± 0.188
  \\
   & Att-UNet \cite{attunetcode, oktay2018attention} &2018&80.05 ± 0.079
 &  87.62 ± 0.078
  & 78.38 ± 0.151

& 87.88 ± 0.095

 \\

    &   MedT \cite{valanarasu2021medical,medtcode}&2021& 80.34 ± 0.034
 &  87.77 ± 0.107
  & 79.29 ±  0.411
& 88.45 ± 0.251
    \\
  & ConvUNeXt \cite{han2022convunext, convunextcode} &2022&80.51 ± 0.043
 &  87.99 ± 0.049
  & 78.71 ± 0.128
& 88.09 ± 0.080
  \\
 &  UNeXt-S \cite{valanarasu2022unext,unextcode}&2022&80.70  ±   0.226  &  88.17 ±  0.194  & 79.26 ±  0.497 & 88.43 ± 0.309   \\
 & MALUNet \cite{ruan2022malunet,malunetcode}&2022&80.95 ± 0.393 &  88.25 ± 0.315   & 79.99 ± 0.644 & 88.88 ± 0.398 \\

  &    EGE-UNet \cite{ruan2023ege,egeunetcode}&2023& 81.16 ± 0.104&88.36 ± 0.086
 &80.28 ± 0.363
  &89.06 ± 0.223 \\

 &    UCM-Net (ours) &2023& 81.26 ± 0.030
  &  {88.48 ± 0.109}
  &80.85 ± 0.251   &89.35 ± 0.111
 \\

    \bottomrule
  \end{tabular}
   \begin{tablenotes}
   
      \item $^\ast$: this method needs the pre-train model on training. 
    \end{tablenotes}
 \label{table3}
   \end{threeparttable}
\end{table}

\begin{table}[h]

 \caption{Comparative performance results on models' computations and the number of parameters.}
  \centering
  \renewcommand{\arraystretch}{1.5} 

  \begin{tabular}{l|l|c|c|c|c}
    \toprule

    Models  & Year & Image Size (H x W)& Params↓ &GFLOPs↓&Memory (MB)↓   \\
    \midrule

      TransUNet$^\ast$ \cite{chen2021transunet,transunetcode} & 2021&224 x 224 & 105,277,081

  &24.7278&402.3614  \\

    SANet$^\ast$ \cite{wei2021shallow,sanetcode} & 2021 &256 x 256 & 23,899,497  &5.9983  &  91.9233\\
    SwinNet$^\ast$ \cite{cao2022swin,swinnetcode} & 2021&224 x 224 & 20,076,204

  &5.5635&77.3503 \\
    TransFuse-S$^\ast$ \cite{zhang2021transfuse, transfusecode}& 2021&192 x 256 & 26,248,725  &8.6462&100.8809  \\
    
     \cline{1-6}
    U-Net \cite{ronneberger2015u,unetcode} & 2015  &256 x 256&31,037,633
  &54.7378&119.3992  \\
    Att-UNet \cite{attunetcode, oktay2018attention} & 2018&256 x 256 &34,878,573
  &66.6318&134.0512 \\

       MedT \cite{valanarasu2021medical,medtcode} & 2021 &256 x 256 & 1,564,202
&2.4061&6.9670 \\
   ConvUNeXt \cite{han2022convunext, convunextcode} & 2022&256 x 256 & 3,505,697
 &7.2537&14.3732 \\
   UNeXt-S \cite{valanarasu2022unext,unextcode} & 2022&256 x 256  & 253,561
  &0.1038&1.9673 \\
  MALUNet \cite{ruan2022malunet,malunetcode}& 2022 &256 x 256& 177,943
  &0.0830&1.6788\\
   EGE-UNet \cite{ruan2023ege,egeunetcode}& 2023&256 x 256  &53,374
  &0.0721&1.2036
 \\
   UCM-Net(ours) &  {2023}&256 x 256  & 49,932
  &0.0465 &1.1905
  \\

    \bottomrule
  \end{tabular}

\label{table2}
   \begin{tablenotes}
   
      \item $^\ast$: this method needs the pre-train model on training. 
    \end{tablenotes}
\end{table}

\begin{figure}[h]
  \centering
  \includegraphics[width=1\linewidth]{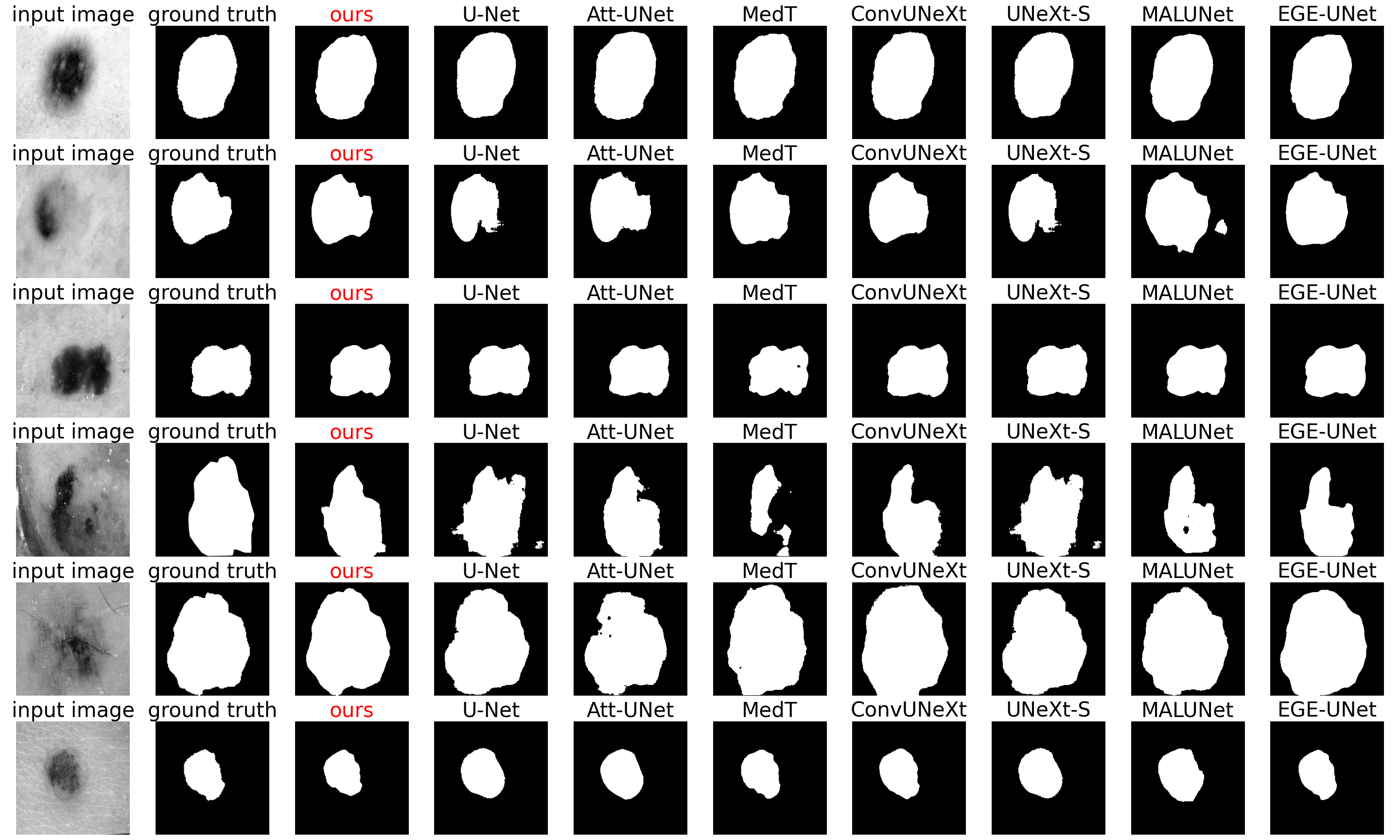}
  \caption{Vision performance comparison on samples}
  \label{vision}
\end{figure}
 \textcolor{black}{Table \ref{table1}-\ref{table3} comprehensively evaluates the performance of our UCM-Net, a novel skin lesion segmentation model, compared to well-established models, using the widely recognized PH2, ISIC2017 and ISIC2018 datasets. Introduced in 2023, UCM-Net is a robust and highly competitive solution in this domain.}
One of the key takeaways from the table is UCM-Net's ability to outperform EGE-UNet, which had previously held the title of the state-of-the-art model for skin lesion segmentation. Our model achieves superior results across various prediction metrics, emphasizing its advancement in the field and its potential to redefine the standard for accurate skin lesion delineation.
Moreover, UCM-Net's performance is notably competitive even when compared to SwinNet, a model that relies on pre-trained models during training. Table \ref{table2} complements this assessment by comparing computational aspects and the number of parameters for various segmentation models. Remarkably, UCM-Net, operating with the same number of channels \{8,16,24,32,48,64\} and image size, as EGE-UNet, boasts fewer parameters and lower GFLOPs. Additionally, even when compared to TransUNet, TransFuse-S and SwinNet, which operate with smaller image sizes, UCM-Net demonstrates faster computational speed. In Figure \ref{vision}, we present a visual exhibition
of all the non-pretrained models' segmentation outputs. This figure directly compares our segmentation results, those produced by other methods, and the ground truth, all displayed side by side using representative sample images. Notably, our segmentation results demonstrate a remarkable level of accuracy, closely resembling the ground truth annotations. Tables \ref{table1}-\ref{table3} and Figure \ref{vision} collectively underscore UCM-Net's exceptional performance and efficiency in skin lesion segmentation, affirming its potential to make a substantial impact in advancing early skin cancer diagnosis and treatment.

\subsection{Ablation results} 

\begin{figure}[!h]
  \centering
  \includegraphics[width=0.85\linewidth]{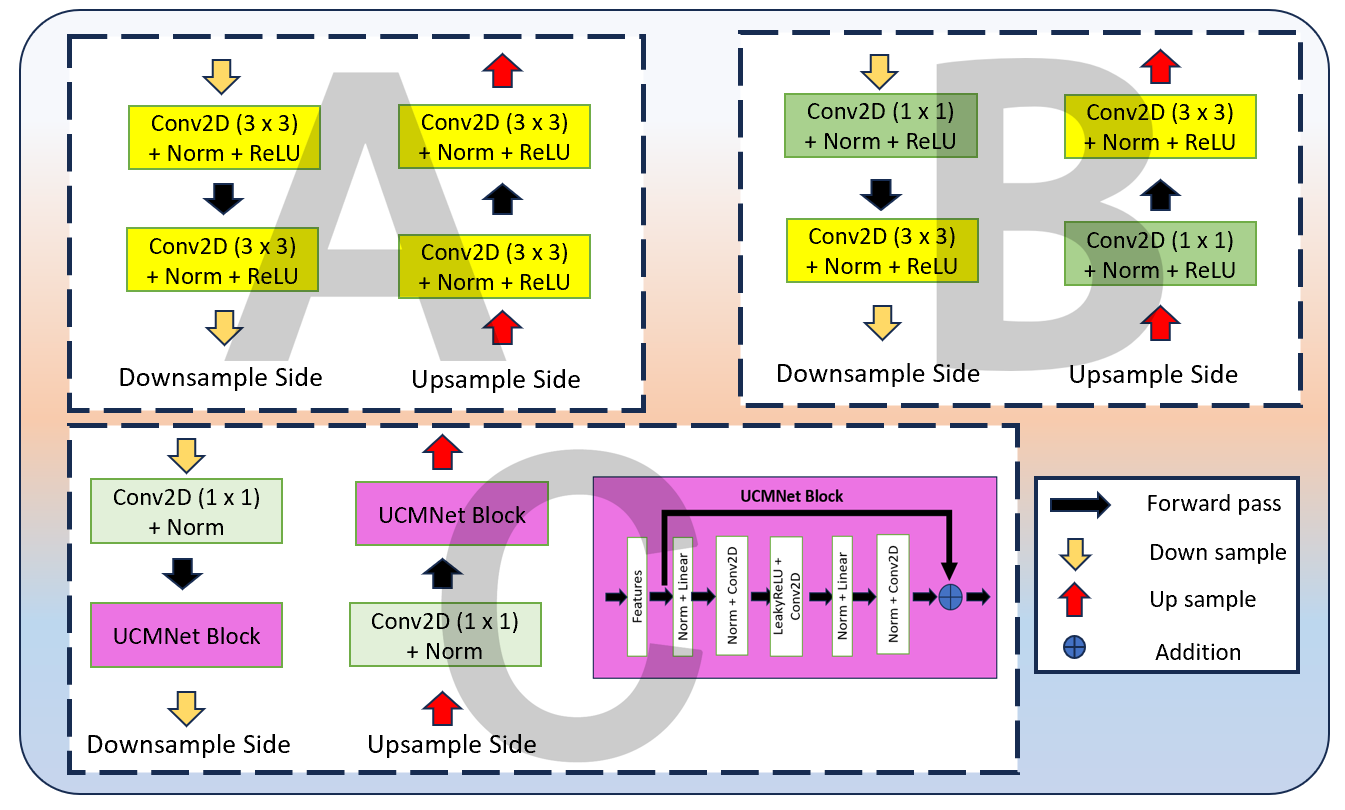}
  \caption{Stage Block Structures in ablation experiments.}{ A: U-Net and U-Net Variant. B: U-Net Variant 1. C: UCM-Net}

  \label{Stages}
\end{figure}

To demonstrate the efficiency and effectiveness of our proposed modules, we conducted a series of ablation experiments on dataset ISIC2017. We develop UCM-Net based on U-Net. Figure \ref{Stages} shows the different block structures in the stages among the compared models. Table \ref{tableforablation} shows the ablation experiments' results including the number of parameters, Giga Flops, mIoU score and mDice score. U-Net variant and variant 1 are six-stage U-Nets with the stage channels \{8,16,24,32,48,64\}. The details of UCM-Net are illustrated in Figure \ref{UCM-Net Structure}.

\begin{table}[!h]
  \centering
 \begin{threeparttable}
 \caption{Ablation experiments' results on the ISIC2017 dataset}

  \renewcommand{\arraystretch}{1.5} 

  \begin{tabular}{l|c|cc|cc}
    \toprule
   
    Models &Structure Reference&Params↓ &GFLOPs↓      &mIoU(\%)↑ &mDice(\%)↑  \\
    \midrule
U-Net(baseline)&Figure \ref{Stages} (A)&310,376,33&54.7378&78.34&86.23\\
U-Net Variant&Figure \ref{Stages} (A)&248,531&0.5715&78.48&86.22\\
U-Net Variant 1&Figure \ref{Stages} (B)&148,157&0.3700&73.89&82.36\\
UCM-Net&Figure \ref{Stages} (C)&49,932&0.0465&79.76&86.94\\
UCM-Net + Group Loss&Figure \ref{Stages} (C)&49,932&0.0465&80.63&87.64\\
UCM-Net + Group Loss(ours)&Figure \ref{Stages} (C)&49,932&0.0465&81.01&87.98\\

    \bottomrule
  \end{tabular}

  \label{tableforablation}
   \end{threeparttable}
\end{table}
\begin{figure}[!h]
  \centering
  \includegraphics[width=1\linewidth]{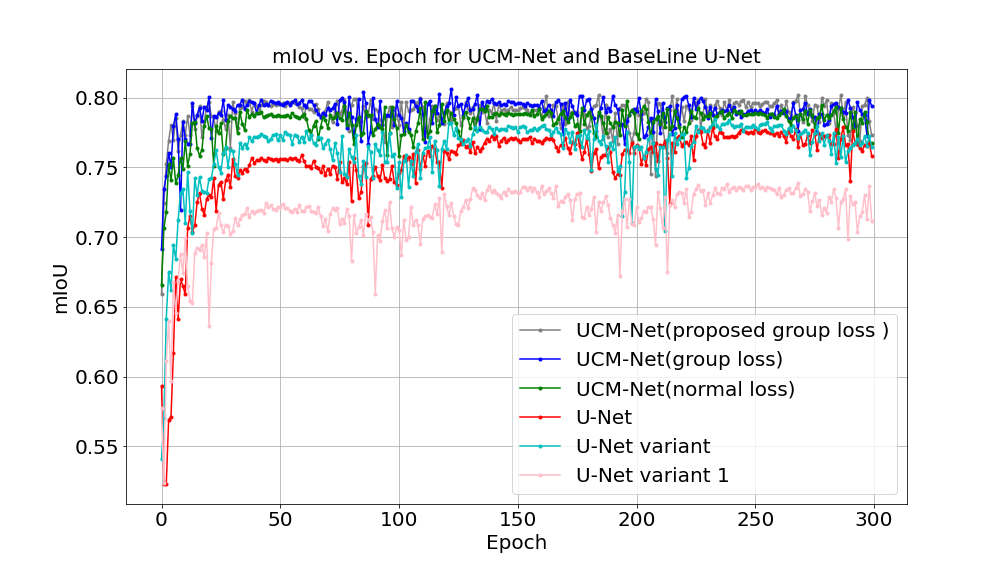}
  \caption{IoU vs Epoch results of ablation experiments}
  \label{his_acc}
\end{figure}
\begin{figure}[!h]
  \centering
  \includegraphics[width=1\linewidth]{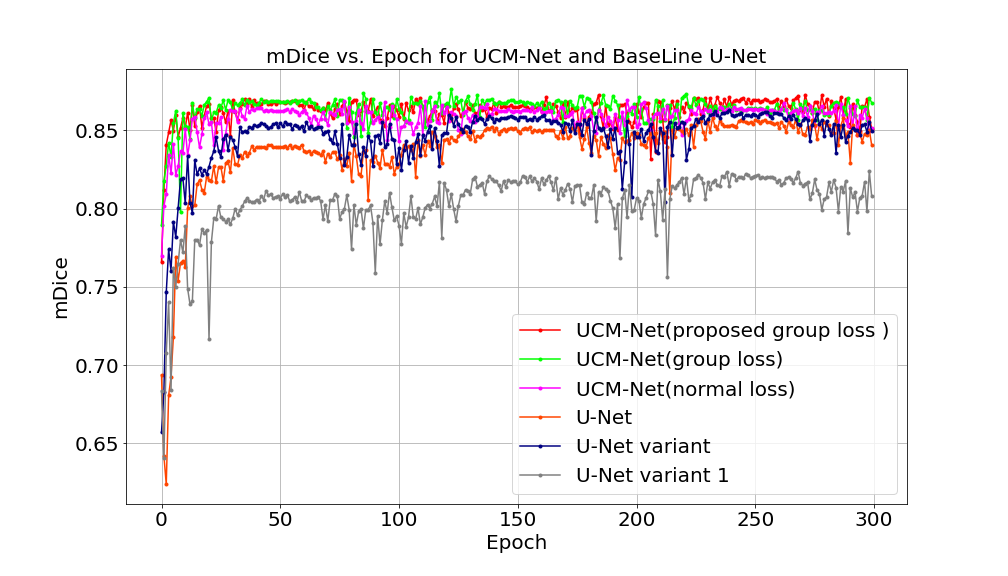}
  \caption{Dice vs Epoch results of ablation experiments}
  \label{his_dice}
\end{figure}
As the results of the U-Net and U-Net variants are shown in Table \ref{tableforablation}, although the number of parameters is reduced, the U-Net variant performs better with the six-stage structure. When we set the one convolution kernel to 1 to reduce the number of parameters, the model's performance drops severely. However, when we replaced the convolution with our proposed UCMNet block, the results showed that the model's performance improved significantly. The UCM-Net, as depicted in Figure \ref{UCM-Net Structure} and Figure \ref{Stages}(C), with 49,932 parameters and 0.0465 GFLOPs, outperforms the U-Net variant with 248,531 parameters and the baseline U-Net with 31,037,633 parameters in terms of both mean Intersection over Union (mIoU) and mean Dice Similarity Coefficient (mDice) metrics.

 \textcolor{black}{Furthermore, when incorporating the proposed Group Loss into the UCM-Net architecture, denoted as "UCM-Net + Group Loss(ours)", the model's performance continued to excel. This enhancement resulted in a higher mIoU of 81.00\% and a mDice of 87.98\%, demonstrating the effectiveness of the proposed Group Loss in further improving segmentation accuracy. Figures \ref{his_acc}-\ref{his_dice} show that "UCM-Net + Group Loss(ours)" always presents the high scores of mIoU and mDice with the training epoch increase. 
 }

The above findings in the ablation experiments underscore the significance of architectural innovations such as the UCMNet block in achieving superior semantic segmentation performance, even with fewer parameters and computational complexity than the U-Net baseline. We note the batch size will affect the prediction performance. \textcolor{black}{So, we tested our model with different batch sizes and reported it in Table \ref{table6} in the appendix. }

\section{Conclusion}
This paper introduces UCM-Net, a novel, lightweight, and highly efficient solution. UCM-Net combines MLP and CNN, providing robust feature learning capabilities while maintaining a minimal parameter count and reduced computational demand. We applied this innovative approach to the challenging task of skin lesion segmentation, conducting comprehensive experiments with a range of evaluation metrics to showcase its effectiveness and efficiency. The results of our extensive experiments unequivocally demonstrate UCM-Net's superior performance compared to the state-of-the-art EGE-UNet. Remarkably, UCM-Net is the first model with fewer than 50KB parameters and consuming less than 0.05 GLOPs for skin lesion segmentation. \textcolor{black}{
Looking forward to future research endeavors, we aim to expand the application of UCM-Net to other critical medical image tasks, advancing the field and exploring how this efficient architecture can contribute to a broader spectrum of healthcare applications. This potential revolution in utilizing deep learning for medical image analysis opens up numerous possibilities for enhancing patient care and diagnostic accuracy. However, it is essential to acknowledge that compared to models leveraging pre-trained components, UCM-Net currently demonstrates lower accuracy performance. Recognizing this limitation, our future efforts will focus on optimizing UCM-Net to improve its accuracy. Such advancements are crucial for ensuring that the model maintains its efficiency and competes favorably in performance with existing state-of-the-art solutions. By addressing these challenges, we aim to advance the field further and expand the impact of deep learning in healthcare applications, making significant contributions to medical imaging and beyond.\\
}

\bibliographystyle{unsrt}  
\bibliography{references}  
\newpage
\section{Appendix}

\begin{table}[!h]
  \centering
 \begin{threeparttable}
 \caption{Batch size experiments' results on the PH2 dataset}

  \renewcommand{\arraystretch}{1.5} 

  \begin{tabular}{l|c|cc|cc}
    \toprule
   
    Models &Batch Size&mIoU(\%)↑ &mDice(\%)↑    &mIoU*(\%)↑ &mDice*(\%)↑   \\
    \midrule
  TransUnet$^\ast$ \cite{chen2021transunet,transunetcode} &8&90.77 ± 0.030
 & 95.09 ± 0.014
  & 91.89 ± 0.148
& 95.77 ± 0.080
  \\
    SANet$^\ast$ \cite{wei2021shallow,sanetcode} &8&89.99 ± 0.349
 &  94.62 ± 0.206

  & 91.06 ± 0.481
&95.32 ± 0.264   \\

    SwinNet$^\ast$ \cite{cao2022swin,swinnetcode} &8&88.92 ± 0.493
 & 94.03 ± 0.289
  & 89.96 ± 0.388
& 94.71 ± 0.215
  \\
   TransFuse-S$^\ast$ \cite{zhang2021transfuse, transfusecode}&8&91.15 ± 0.131 & 95.30 ± 0.078 
  &92.16 ± 0.113 &95.92 ± 0.061   \\
    
    \cline{1-6}
    \multirow{ 6}{*}{UCM-Net + Group Loss(ours)}
&1&89.68 &94.45&89.96&94.72\\
&2& 89.88&94.57&90.39&94.95\\
&4& 89.57&94.38&90.18&94.84\\
&8& 89.39&94.28&89.67&94.55\\
&16& 88.44&93.74&88.85&94.10\\
&32& 89.15&94.15&90.23&94.86\\

    \bottomrule
  \end{tabular}

  \label{table6}
   \end{threeparttable}
\end{table}
 We tested our model with different batch sizes and reported it in Table \ref{table6} in the appendix.

\end{document}